\documentstyle[preprint,aps,epsf]{revtex}
\begin{document}
\baselineskip=18pt
\def\be{\begin{equation}}
\def\ee{\end{equation}}
\def\bear{\begin{eqnarray}}
\def\eear{\end{eqnarray}}
\def\E{{\rm e}}
\def\bearst{\begin{eqnarray*}}
\def\eearst{\end{eqnarray*}}
\def\peleven{\parbox{11cm}}
\def\peffec{\peight{\bearst\eearst}\hfill\peleven}
\def\pspace{\peight{\bearst\eearst}\hfill}
\def\ptwelve{\parbox{12cm}}
\def\peight{\parbox{8mm}}
\title
{Evolution of Scale-invariant Inhomogeneities in Standard Cosmology}
\author
{Elcio Abdalla$^a$ and 
Roya Mohayaee$^b$}
\address
{\it Instituto de 
F\'\i sica-USP, C.P. 66.318, S\~ao Paulo, Brazil,\\
$^a$eabdalla@fma.if.usp.br\\
$^b$roya@fma.if.usp.br\\
}
\date{14/11/97}
\maketitle

%%%%%%%%%%%%%%%%%%%%%%%%%%%%%%%%%%%%%%%%%%%%%%%%%%%%%%
%ABSTRACT
%%%%%%%%%%%%%%%%%%%%%%%%%%%%%%%%%%%%%%%%%%%%%%%%%%%%%%
\begin{abstract}

It is shown that density fluctuations obey a scaling law 
in an open Friedmann universe. In a flat universe, the
fluctuations are not scale-invariant. We compute the 
growth rate of adiabatic scale-invariant density fluctuations 
in flat, open and inflationary universes. We find that, 
given a sufficiently long time, the density perturbations 
decay away in the Einstein-de-Sitter universe. On the 
contrary, the rapid growth of the density instabilities 
makes an open universe inhomogeneous in a time scale 
comparable to the age of our universe. We also find that 
the fluctuations grow exponentially in a flat inflationary 
universe.

\end{abstract}

\newpage

%%%%%%%%%%%%%%%%%%%%%%%%%%%%%%%%%%%%%%%%%%%%%%%%%%%
\section{Introduction}
%%%%%%%%%%%%%%%%%%%%%%%%%%%%%%%%%%%%%%%%%%%%%%%%%%%
\indent

Friedmann-Robertson-Walker solution of the Einstein field equation
describes a homogeneous and isotropic universe and forms the basis of
standard cosmology. The strongest support for this model is provided
by the cosmic microwave background radiation which is
isotropic to about a part in $10^4$ on angular scale from $10^{''}$ to
$180^o$. However, the model also faces many open questions among which are 
the horizon problem, the flatness problem 
and the absence of antibaryons and monopoles. 

With the availability of more and more reliable three-dimensional catalogues 
in the last few years, the standard cosmology has been
confronting new challenging problems \cite{voids}. 
The observations of the red shift surveys indicate 
the existence of voids and filaments up to large
scales in the universe \cite{voids} and raise doubts about the hypothesis of
homogeneity, which is at the foundation of the Friedmann cosmology. 
The precise correlation length at which the matter 
distribution, given in these
catalogues, smoothes down and becomes homogeneous is the
subject of strong controversy \cite{peeble,pietronero,schramm,davis}. 
It is, however, unanimously agreed that
the observed inhomogeneities point towards a scale-invariant
distribution at least up to a certain scale \footnote{This scale ranges
from $5$ Mpc, the lowest set by the supporters of the standard cosmology,
\cite{peeble,davis} to $1000$ Mpc, the highest claimed 
by its critics \cite{pietronero}.}.
Since the Friedmann cosmology is inappropriate for describing a
non-homogeneous universe, we seek for more general solutions 
of the Einstein field equation.

The general solution of the Einstein equation, not requiring the
homogeneity assumption, was given by Tolman for spherically
symmetric dust flows in comoving coordinates \cite{tolman}. 
Starting from the Tolman
solution, it was shown that there exist spherically
symmetric inhomogeneous cosmological models which approach 
Friedmann universe at sufficiently large times, but whose initial 
density is an arbitrary function of the radial variable \cite{bonnor}.
In more recent works, Tolman solution was used to model a hierarchical
(fractal) universe which would accommodate the results of the red shift 
surveys \cite{ribeiro}. 

On the non-relativistic front, advances have been made in 
describing the non-homogeneous distribution of matter 
at small scales \cite{turbulence}. 
Recently, it has been shown that the observed inhomogeneous
matter distribution in the interstellar
medium can be explained on the basis of Newtonian 
self-gravity by a purely field theoretical model \cite{devega}. 
Using this model, 
it was further shown
that, at small scales, up to 100 pc, the density-density correlator of the 
interstellar medium obeys a scaling law \cite{devega}.

At higher scales, similar analytical works on the 
dynamics of an inhomogeneous 
matter distribution in the universe, are rather scarce 
\cite{bonnor,ribeiro}. A theoretical
framework explaining the origin of 
the inhomogeneity of the galaxy and cluster 
distributions, observed in the three-dimensional catalogues, is not
yet available.

In this work, we employ a perturbative method to explain how a small
scale-invariant fluctuation, seeded in a homogeneous background, evolves into 
an inhomogeneous universe.
We start, not with the Tolman dust
solution, but with a general cosmological model which
allows non-vanishing pressure. 
We obtain our cosmological model by perturbing
the Friedmann scale factor and allowing it to be 
radial-coordinate dependent through a general fluctuation term.
We compute this fluctuation by assuming that the universe consists of a
gas of point particles which interact through Einstein gravity. 
Using the grand canonical ensemble of these particles, 
we show that, at all scales, 
as for the interstellar
medium, the density-density correlator obeys a scaling law in
an open universe. We also show that in a flat universe,
the correlator does not scale. 

Having shown that the
fluctuation obeys a scaling law, we subsequently perturb the
energy-momentum tensor by a scale-invariant adiabatic perturbation 
and reduce the Einstein field
equations to a second-order hypergeometric equation.
The hypergeometric solutions
are used to obtain the ratio of the perturbation to the
background homogeneous density. 

We evaluate the growth rate of the
perturbation during the matter and radiation-dominated eras up to the
present time. We show that, in the lifetime of our universe, 
any scale-invariant adiabatic inhomogeneity seeded in the
open Friedmann universe grows to dominate over the homogeneous
background. On the contrary, in a flat universe, 
an adiabatic scale-invariant perturbation decays away in this time scale. 
These results are in complete agreement with those obtained for the 
Tolman dust solution \cite{ribeiro}. We extend our results to 
the inflationary universe
and show that, given a sufficiently large time, 
the exponential growth of the adiabatic perturbations makes the
matter distribution inhomogeneous. Finally, we obtain the
thermodynamical conditions required for the growth of fluctuation in 
a universe with an arbitrary pressure.

The plan of this article is as follows. We start section II with
Einstein lagrangian and derive the
density-density correlator for a perturbed Friedmann universe. 
In Section III, we perturb
the Einstein field equation and the energy-momentum conservation
equation and obtain a second-order hypergeometric differential
equation. In
section IV, we solve this equation for a flat universe and show that
the perturbation decays in a time scale comparable 
to the age of our universe. In Section V, we show that the
scale-invariant fluctuation grows exponentially in an inflationary universe. 
In Section VI, we discuss the hypergeometric
solution of the Einstein equation for a 
non-flat universe with an arbitrary pressure.
The consequences of interchanging the polar with
the radial pressure in solving the Einstein field equation is discussed
in Section VII.
Section VIII is devoted to the conclusion.

%%%%%%%%%%%%%%%%%%%%%%%%%%%%%%%%%%%%%%%%%%%%%%%%%%%%%
\section{Scaling behaviour of the density function}
%%%%%%%%%%%%%%%%%%%%%%%%%%%%%%%%%%%%%%%%%%%%%%%%%%%%%
\indent

In this section, we evaluate the density-density correlator for flat
and open universes. We use the field theoretical method developed in
ref.\cite{devega} to obtain the infrared, {\it i.e.}\ large distance,
correlator from the grand partition
function of a gas of particles interacting through Einstein
gravitational potential.

The dynamics of our universe is described by
the Einstein lagrangian,
\be
{\cal L}=\sqrt{-g}\left( g^{\mu\nu} T_{\mu\nu}-{1\over 16\pi G}
R\right).
\label{lagrangian}
\ee
A homogeneous and isotropic universe is obtained by using
the Friedmann-Robertson-Walker metric
\be
{\rm diag}(g_{\mu\nu})=\left(-1,{R_0^2\over 1-kr^2},R_0^2r^2
,R_0^2r^2{\rm sin}^2\theta\right) \label{metric}
\ee
and a diagonal energy-momentum tensor
$T_{\mu\nu}$ of a perfect fluid in the lagrangian (\ref{lagrangian}).

We perturb the scale factor $R_0$ by a small perturbation term $\delta
R$. The fluctuation in the Einstein lagrangian, for a matter-dominated
universe is
\footnote{
For the purpose of evaluating the density correlator, it suffices
to consider a static problem where all the time derivatives are set to
zero.},
\be
\delta{\cal L}=\delta R O \delta R + {\rho_1 r^2\over
\sqrt{1-kr^2}}\delta R
\ee
where
\begin{eqnarray}
O&=&{\partial\over\partial r}\left({r^2\sqrt{1-kr^2}\over 8\pi G
    R_0}{\partial\over\partial r}\right)+
    {3R_0^2 r^2\rho\over \sqrt{1-kr^2}},\\
\rho_1&=& 3 R_0^2\rho + {3k\over 8\pi G}.\label{rho1}
\end{eqnarray}
Our aim is to find the two body hamiltonian which describes the
interaction of the point objects in the universe. We shall neglect all
higher-order interactions and only work with the quadratic terms.
We also keep distances large.
To obtain an operator for the two-body interaction, we substitute the
equation of motion of the fluctuation $\delta R$ back into the
lagrangian and obtain the effective lagrangian \footnote{
Since we are considering a static problem , the hamiltonian 
, $H=P_R \dot R -{\cal L}$, only differs from the lagrangian by a
minus sign.} 
\be
{\cal L}_{\rm eff}=-\rho_1\, A^{-1}\,\rho_1 , \label{ll}
\ee
where the interaction operator is
\be
A^{-1}={r^2\over 2\sqrt{1-kr^2}}O^{-1}
{r^2\over 2\sqrt{1-kr^2}}. \label{a}
\ee

Next, we write the grand canonical
ensemble for the universe. We consider the universe to be
composed of a gas of particles of 
mass $m$ interacting through the gravitational field, as given by the
two body lagrangian, (\ref{ll}), and the interaction 
operator, (\ref{a}), at an equilibrium
temperature $T$. The grand partition function, allowing for the variation
of the number of particles $N$, is
\be
{\cal Z}=\sum_{N=0}^\infty {z^N\over N^!}\int\cdots\int\prod_{l=1}^N
{dp_l\over 2\pi}dr_l e^{-H_N/ T},
\ee
where $z$ is the fugacity and the Hamiltonian $H_N$ is
\be
H_N=\sum_{l=1}^N{p_l^2\over 2m} +\int dr dr^\prime 
\left(\rho_1(r) A^{-1} \rho_1(r^\prime)\right).
\ee
Inserting this in the partition function and integrating over the
momenta, we obtain
\be
{\cal Z}=\sum_{N=0}^\infty{1\over N !}\left(z\sqrt{{mT\over 2\pi}}\right)^N
\prod_{l=1}^N\int dr_l \exp\left[{{-1\over T}\int \rho_1(r)
  A^{-1}\rho_1(r^\prime)} dr dr^\prime \right].\label{partition0}
\ee
The last integral can be expressed as a functional integral, {\it i.e.},
\be
 \exp\left[{-1\over T}\int \rho_1(r)
  A^{-1}\rho_1(r^\prime) dr dr^\prime\right]=\int {\cal D}\zeta 
e^{-\int dr( -\zeta A \zeta+2\rho_1\zeta/\sqrt T)}.\label{partition1}
\ee
At this point we define the density as,
\be
\rho_1=\sum_{l=1}^N\delta(r-r_l)
\ee
so that the partition function can be rewritten as  
\footnote{Note that in the exponential term the action is
unbounded from below, which is due to the attractiveness of the
gravitational interaction. The divergence can be easily eliminated by
a short distance cut-off. 
For a comprehensive discussion see ref. \cite{devega}} ,
\be
{\cal Z}
=\int{\cal D}\zeta e^{-\int dr\left(-\zeta A\zeta
-z\sqrt{mT/2\pi}e^{-2\zeta/\sqrt T}\right)} ,\label{partition2}
\ee
which describes an exponential self-interaction for the $\zeta$
field.

The equations of motion, obtained from the expressions
(\ref{partition0}), (\ref{partition1}) and (\ref{partition2}) are
\be
{\partial{\cal Z}\over\partial\zeta}=\langle 2 A\zeta
-2{\rho_1\over\sqrt T}\rangle=
\langle 2 A\zeta -2z\sqrt{m\over 2\pi} e^{-2\zeta/\sqrt T}\rangle=0,
\ee
where the angled brackets stand for the average values.  
Thus, the average density is given by,
\be
\langle \rho_1(r) \rangle = z\sqrt{mT\over 2\pi}\langle
e^{-2\zeta/\sqrt T}\rangle.
\ee  
This expression can also be obtained by introducing an external source
for $\rho_1$ in (\ref{partition1}). Analogous procedure leads to the
two-point function 
\be
\langle \rho_1(r) \, \rho_1(r^\prime)\,\rangle = 
{z^2 mT\over 2\pi}\langle e^{-2\zeta/\sqrt T} 
e^{-2\zeta/\sqrt T}\rangle 
={z^2 mT\over 2\pi}
e^{{4\over T}\langle \zeta (r) \, \zeta(r^\prime)\rangle},\label{correlator}
\ee
where the last step is valid because the field $\zeta$ is
one-dimensional in this time-independent problem.

The Green function,
\be
{\cal G}=\langle \zeta (r)\,\, \zeta(r^\prime)\rangle =-A^{-1},
\ee
can be computed
for both flat and open universes in the limit of large radial distances.

For a flat universe, $k=0$, the infrared Green function 
satisfies the differential equation,
\be
{\partial\over\partial r}
\left({1\over r^2}
{\partial {\cal G}\over\partial r}\right)=\delta(r-r^\prime)
\ee
and is given by
\be
{\cal G}={\epsilon(r-r^\prime)\over 6}\left[r^3-{r^\prime}^3\right]=
{1\over 6}\vert r^3-{r^\prime}^3\vert
\ee
where $\epsilon(r-r^\prime)$ is the step function.
Substituting this Green function together with 
the relation (\ref{rho1}) in (\ref{correlator}), 
we obtain the density-density correlator
\be
\langle \rho (r)\, \rho(r^\prime)\rangle_{k=0}\sim \exp\left[{-4\pi G
R_0(t)\over 3T}\vert r^3-{r^\prime}^3\vert\right]\, ,
\ee
which shows that for a flat universe the correlator does not scale and
therefore, the assumption of homogeneity is valid. 

For an open universe, the Green function obeys the differential equation
\be
{\partial\over\partial r}\left(r{\partial {\cal G}\over\partial r}\right)=
\delta(r-r^\prime)
\ee
leading to
\be
\langle \zeta(r) \zeta(r^\prime) \rangle_{k=-1} =
\epsilon(r-r^\prime) {\rm ln}\left({r\over r^\prime}\right),
\ee
which on substitution in (\ref{correlator}) 
and using (\ref{rho1}) yields
\be
\langle \rho(r) \rho(r^\prime) \rangle _{k=-1}\sim
{\left\vert {r^\prime\over r}\right\vert^{8\pi G\epsilon R_0/T}}\,\,.
\ee
Thus, in an open universe, for large distances,
the density correlator scales and
the assumption of the homogeneity of the universe has to be revised.

%%%%%%%%%%%%%%%%%%%%%%%%%%%%%%%%%%%%%%%%%%%%%
\section{Einstein equations}
%%%%%%%%%%%%%%%%%%%%%%%%%%%%%%%%%%%%%%%%%%%%%
\indent

In the previous section, we have perturbed the Friedmann scale factor
by a general fluctuation term and have shown that in an open universe the
fluctuation obeys a scaling law. In this section, we study the
dynamics of scale-invariant fluctuations. We perturb the
the energy-momentum tensor and Einstein field equation around 
the background homogeneous Friedmann universe by scale-invariant
adiabatic fluctuations. We show that Einstein equations reduce to a
second-order hypergeometric equation. Subsequently, the
density fluctuations can be obtained from the hypergeometric solutions.

The Einstein field equation corresponding 
to the lagrangian (\ref{lagrangian}) is
\be
R_{\mu\nu}-{1\over 2} g_{\mu\nu}R=-8\pi G T_{\mu\nu}.\label{fieldeqn}
\ee

We expand the scale factor, density and pressure
around the background homogeneous functions 
$R_0$, $\rho_0$ and $P_0$ respectively , {\it i.e.},
\bear
\qquad\qquad R(r,t)&=&R_0(t)+X(t)r^{-\sigma},\label{reqn}\\
g_{tt}T^{tt}=
\rho(t,r)&=&\rho_0(t) + \lambda(t)r^{-\gamma},\label{rhoeqn}\\
g_{rr}T^{rr}=P_r(t,r)&=&P_0(t) + \pi(t)r^{-\gamma},\label{preqn}\\
g_{\theta\theta}T^{\theta\theta}=
P_\theta(t,r)&=&P_0(t) + \pi_\theta (t)r^{-\gamma}.\label{pthetaeqn}
\eear
where the fractal codimension $\gamma$ is approximately 1
\cite{reportnero} and
the comoving radial coordinate has been used; {\it i.e.}, 
$X(t)=q(t)R_0^{-\gamma}(t)$ {\it etc.}.

By expanding the energy-momentum conservation equations,
\bear
{\partial\over\partial t}\left(R^3\rho\right)-\dot R
R^2(R_r+2P_\theta)+{1-kr^2\over r^2 }{\partial\over\partial r}
\left(T_{r0}r^2R\right)-krRT_{r0}&=&0\label{energy2}\\
{\partial\over\partial t}\left(R^3T_{ro}\right)-\frac R
r^2{\partial\over\partial r}\left(r^2R^2P_r\right)+{P_\theta R\over r^2}
{\partial\over\partial r}\left(R^2r^2\right)&=&0.\label{energy1}
\eear
around the homogeneous background, we find
that to first-order in power of $\gamma$ and for very large $r$ 
\footnote{The off-diagonal component of
energy-momentum tensor is expected to arise for all types of
inhomogeneous universes, including the Tolman universe\cite{ribeiro}. 
This component is also expanded as, 
$T_{0r}(t,r)=\tau(t)r^{-\gamma^\prime-1}$.} :
\be
\gamma=\sigma=\gamma^\prime.
\ee

This is substituted in the expressions (\ref{reqn}-\ref{pthetaeqn}) 
which are subsequently used in the Einstein field equation (\ref{fieldeqn})
to yield
\bear
6\frac{\dot R_0}{R_0}\left( \frac{\dot X}{R_0}-X\frac{\dot R_0}{R_0}\right)
+2k\frac{\dot X}{R_0^3} \left(\gamma +1\right) \left(\gamma -3\right)
&=& -8\pi G\lambda(t),\label{x1}\\
\frac{\ddot X}{R_0}- X\frac{\ddot R_0}{R_0^2}-X\frac{\dot R_0^2}{R_0^3}
+\dot X \frac{\dot R_0}{R_0^2}-k\left(\gamma +1\right)\frac{X}{R_0^3}
&=& -4\pi G\pi_r(t),\label{x2}\\
\frac{\dot R_0}{R_0}\left( \frac{\dot X}{R_0}-X\frac{\dot R_0}{R_0}\right)
+\frac{\ddot X}{R_0}- X\frac{\ddot R_0}{R_0^2} +\frac
{k (\gamma +1)(\gamma -2)}{2R_0^3} &=& -4\pi G\pi_\theta(t).\label{x3}
\eear
 
A linear relationship between the homogeneous pressure $P_0$
and the density $\rho_0$, {\it i.e.}\ $P_0=\omega\rho_0$, is usually
taken as the homogeneous equation of state \cite{weinberg,kolb}. It is most
practical to use an adiabatic perturbation where
$\pi_r =\alpha\lambda$
and the proportionality constant $\alpha$ takes a value
very close to that of $\omega$. The coupled equations
(\ref{x1},~\ref{x2}) now reduce to the differential equation, 
\be
\ddot X-\left((3\alpha-1)\frac{\dot R_0} R_0\right)\dot
X+\left((3\alpha-1){\dot R_0^2\over R_0^2}-{\ddot R_0\over
R_0}-{k(1+\gamma)(1+\alpha\gamma-3\alpha)\over R_0^2}\right)X=0
\ee
The first-derivative term is eliminated, by the redefinition
\be
X=R_0^{3\alpha-1\over 2} Y, \label{xyeqn}
\ee
to yield the second-order differential equation
\be
\ddot Y+\left(6\pi
G(\alpha-1)(\alpha\rho_0-P_0)-A_3\frac k {R_0^2}\right) Y=0 ,\label{mastereqn}
\ee
where
\be
A_3=
\left((1+\gamma)
(1+\alpha(\gamma-3)\right)-\frac
34(\alpha-1)(3\alpha-1).\label{a3}
\ee

In the remaining part of this work, we solve this differential equation
for different types of universes in different
thermodynamical setups. From the solutions, 
the perturbation coefficient $X(t)$
and subsequently, by using the Einstein equation
(\ref{x1}), the coefficient of the density 
fluctuation $\lambda(t)$ can be found. Using these results, we
establish whether, in the lifetime of our universe,
the fluctuation grows to overrule the
homogeneous background or else is overruled by it
and the universe remains homogeneous.

%%%%%%%%%%%%%%%%%%%%%%%%%%%%%%%%
\section{Flat universe}
%%%%%%%%%%%%%%%%%%%%%%%%%%%%%%%%
\indent

In this section, we solve the hypergeometric equation
(\ref{mastereqn}) for Einstein-de-Sitter universe. 
We show that, given a sufficiently long time,
any scale-invariant adiabatic fluctuation decays away in a 
flat universe. This confirms the result of Section II: that the
density fluctuation does not scale in a flat universe. 

For a flat homogeneous universe, the density 
function is given by\cite{kolb}
\be
\rho_0\sim R_0^{-3(1+\omega)}\qquad;\qquad
R_0\sim t^{2\over 3(1+\omega)}
\ee
Substituting these in the differential equation (\ref{mastereqn}) gives
\be
\ddot Y-{(\alpha-1)(\alpha-\omega)\over(1+\omega)^2} {Y\over t^2}=0
\ee 
which can be solved exactly to yield
\be
X=R_0^{3\alpha-1\over 2}Y=t^{-{3\alpha+1\over 3(1+\omega)}+\frac
12\pm\left\vert{\omega-1-2\alpha\over 2(\omega+1)}\right\vert},\label{asympk=0}
\ee
for the fluctuation coefficient of
the scale factor (see equation (\ref{xyeqn})).
Thus, we find that
$X$ either dominates over $R_0$ or is of the same
order as $R_0$ \footnote{We take the positive
exponent in (\ref{asympk=0}) for large times.}  . The latter is true
when $(\alpha +1)$ and $(\omega +1)$ have the same
sign and $2\vert \alpha +1\vert \ge \vert \omega +1\vert$.
However, since the observations probe the structure of the
matter distribution, it is more appropriate to compare
the density coefficients, $\rho_0$ and $\lambda R_0^\gamma$, than
the scale factors. 
The asymptotic expression for $\lambda$, obtained
by substituting for $X$, (\ref{asympk=0}), in the Einstein equation,
(\ref{x1}), is 
\be
\lambda\sim \cases{t^{(2\alpha-\omega-3)/(1+\omega)}\cr
                   0\cr}
\ee
while for $\rho_0$ is \cite{weinberg,kolb}
\be
\rho_0\sim t^{-2}.
\ee
The growth parameter
\be
\zeta={{\rm ln}\left(\lambda R_0^\gamma/\rho_0\right)\over {\rm ln}t}=
{-2\alpha+\omega-1+{2\over 3}\gamma\over 1+\omega},
\ee
can be used to evaluate the rate of growth of the fluctuation with
respect to the background, given
by, 
\be
\eta=\left({t\over t_0}\right)^\zeta ,  \label{eta}
\ee
where $t_0$ and $t$ mark the beginning and the end of
the thermodynamical era under consideration. Since the growth parameter is
negative for a matter-dominated universe ({\it i.e.},
$\omega=\alpha=0$), the perturbation decays. 
For the Einstein-de-Sitter universe, the perturbation decreases by a
factor of $10^{-2} $, from the beginning of the matter-dominated era
to the present time ({\it i.e.}, for $t/t_0\approx 10^5$).
Thus, the homogeneity hypothesis of the standard cosmology remains
valid in this universe.

%%%%%%%%%%%%%%%%%%%%%%%%%%%%%%%%%%%%%
\section{Inflationary universe}
%%%%%%%%%%%%%%%%%%%%%%%%%%%%%%%%%%%%%
\indent

In this section, we solve the differential equation (\ref{mastereqn})
for an inflationary universe. We show that in all flat inflationary
universes the density fluctuations grow exponentially, leading to an
inhomogneous universe.

At sufficiently large times, all flat, open and closed inflationary
universes are represented by an exponential scale factor 
\be
R_0\sim e^{Ht}
\ee
and a constant matter distribution \footnote{We assume that the Hubble
parameter is constant. However, this is only true for a flat universe.}
\be
\rho_0=-{3H^2\over 8\pi G}\,\, .
\ee 
The coefficient of the $Y$ term 
in the differential equation 
(\ref{mastereqn}) is now constant, leading to the exponential solutions
\be
Y\sim e^{\pm\frac 32(\alpha-1)Ht}
\ee
and, on using (\ref{xyeqn}), to
\be
X\sim\cases{e^{Ht}\cr
e^{(3\alpha-2)Ht}\cr}
\ee
Substituting this result in the Einstein equation (\ref{x1}) we find the
coefficient of the density fluctuation,
\be 
\lambda\sim{-9H^2(\alpha-1)\over 4\pi G} e^{-3(\alpha+1)Ht}.
\ee

Since the homogeneous background
density is constant for an inflationary universe, the perturbation
decays only if
$\alpha >(\gamma/3-1)$. For an adiabatic perturbation,
this range of $\alpha$ is not allowed and therefore 
the fluctuation grows exponentially and 
leads to an inhomogeneous universe.

%%%%%%%%%%%%%%%%%%%%%%%%%%%%%%%%%%%%%%%%%%%%%%%%%%%%%%%
\section{Open universe } 
%%%%%%%%%%%%%%%%%%%%%%%%%%%%%%%%%%%%%%%%%%%%%%%%%%%%%%%%
\indent

In this section, we solve the differential equation (\ref{mastereqn})
for an open universe. We 
show that in an open universe a scale-invariant adiabatic perturbation
grows to dominate over the background homogeneous distribution of
matter. Thus, we confirm the result of
Section II: that the density-density correlator is scale-invariant in an
open universe.

For a non-vanishing $k$, the differential equation (\ref{mastereqn}) is 
complicated. We solve the equation in three steps 
of increasing difficulty. We start with a matter-dominated
universe where the fluctuation makes a zero contribution to the
pressure. We then consider a matter-dominated universe with a very
small pressure caused by fluctuations. Finally, we generalize our
results to include the radiation-dominated universe. 
We derive the thermodynamical conditions necessary for
the growth of perturbation
in a universe with a non-vanishing arbitrary pressure.

%%%%%%%%%%%%%%%%%%%%
\subsection{Matter-dominated open universe\\
$P=0\,\,\,;\,\,\, P_0=0\,\,\, ,\,\,\, \alpha=0$}
%%%%%%%%%%%%%%%%%%%%
\indent

In this case, the homogeneous density and scale factor are given 
parametrically in terms of an angle $\theta$ as
\be
\rho_0\sim R_0^{-3}\quad;\quad
R_0={dt\over d\theta}\quad;\quad
t={\Omega_0\over 2H_0(\Omega_0-1)^{3/2}}(\theta-{\rm sin}\theta)
\label{rreqn}
\ee
where $\theta$ is real for a closed and imaginary for an open
universe and $\Omega$ is the ratio of the present to the
critical density \cite{kolb}. Substituting these in 
the differential equation (\ref{mastereqn}), we obtain
\be
{d\over d\theta}\left({1\over 1-{\rm cos}\theta}{dY\over
    d\theta}\right)-\left(\gamma+\frac 14\right)
{Y\over (1-{\rm cos}\theta)}=0.
\ee
By making the substitution $2z=1-{\rm cos}\theta$, 
we re-write the above equation in the standard
form of a hypergeometric equation,
\be
z(1-z){d^2Y\over dz^2}-\frac 12{dY\over dz}-(\gamma+{1\over 4})Y=0. \label{ha}
\ee
The solutions can be found by using the
Gauss relation for hypergeometric functions
$F(a,b,c;z)$ \cite{morse-feshbach}. That is,
\be
F(a,-1-a,-{1\over 2};z)=2(a+\frac 12)F(a,-a-1,\frac 12;z)-2a
F(a+1,-a-1,\frac 12;z),
\ee
where for our problem, $a=(-1\pm\sqrt{-4\gamma})/2$ 
and $z={\rm sin}^2(\theta/2)$.
The two tabulated solutions of 
(\ref{ha}) are \cite{morse-feshbach},
\bear
Y_{I}&=&{\rm cos}\frac \theta 2{\rm cosh}(\sqrt\gamma\theta)- 
2\sqrt\gamma{\rm sin}\frac\theta 2{\rm sinh}(\sqrt\gamma\theta), \label{y1}\\
Y_{II}&=&\sqrt\gamma {\rm
    sin}\frac\theta 2 {\rm cosh}(\sqrt\gamma\theta)- {\rm
    cos}\frac\theta 2 {\rm sinh}(\sqrt\gamma\theta).\label{y2}
\eear 
Asymptotically, for an open universe, $R_0$ varies linearly with time, 
whereas, using (\ref{xyeqn}) and taking the asymptotic limit of either
(\ref{y1}) or (\ref{y2}) we obtain,
\be
X\sim t^{i\sqrt{\gamma}}.
\ee
Thus, the perturbative corrections to the homogeneous
scale factor are insignificant for large times. On the other hand, 
for the matter distribution, we obtain, by substituting for $X$ in the
Einstein equation (\ref{x1}),
\be
\lambda\sim t^{i\gamma-3}.
\ee
The growth parameter of the density fluctuation
\be
\Re e\,\left({\ln(\lambda R_0^\gamma/\rho_0)\over \ln t}\right)
=\gamma
\ee
shows that , for $\gamma\approx 1$,  
the fluctuation evolves to dominate over the homogeneous background density.
In fact, using expression (\ref{eta}), we find 
that the perturbation grows by a factor of $10^5$ in the matter-dominated
era.
%%%%%%%%%%%%%%%%%%%%%%%%%%%%%%
\subsection{Matter-dominated open universe with a small fluctuation pressure\\
$P_0=0 \, , \, \alpha\not=0$}
%%%%%%%%%%%%%%%%%%%%%%%%%%%%%%
\indent

In this subsection, we study a matter-dominated open universe,
where the fluctuation makes a very small \footnote{ For $P=0$, the
parameter $\alpha$ has to be 
very small to keep the perturbation adiabatic.}, but 
nevertheless nonvanishing, contribution to the pressure.

The homogeneous density and scale factor are the same as
those given in (\ref{rreqn}).
The differential equation (\ref{mastereqn}) can be
written as
\be
{d^2Y\over d\theta^2}-{{\rm sin}\theta\over {1-{\rm
      cos}\theta}}{dY\over d\theta}+\left(-\frac
  92\alpha(\alpha-1){1\over 1-{\rm cos}\theta}-A_3\right)Y=0.
\ee
As in the previous subsection, we make the substitution
 $2z=1-{\rm cos}\theta$ and rewrite the above differential equation as
\be
{d^2Y\over dz^2}-\left({1/2\over z}-{1/2\over z-1}\right){dY\over dz}
-\left({-9\alpha(\alpha-1)\over 4z}-A_3\right)Y=0. \label{last}
\ee
However, unlike the equation (\ref{ha}), this expression needs a further
transformation to become hypergeometric. We make
the substitution
\be
Y=z^\zeta(z-1)^\eta\Psi,
\ee
where $\zeta=3\alpha/2$ or $3(1-\alpha)/2$ and $\mu=0$ or
$\mu=1/2$. Since each of the parameters $\zeta$ and $\eta$ 
takes two values, there are four solutions to the differential
equation (\ref{last}). However, only the
following two solutions :
\be
Y_I=z^{3\alpha/2}F\left({3\alpha\over 2}-\frac
  12+{\sqrt{1-4A_3}\over 2}\, ,\, {3\alpha\over 2}-\frac 12-
{\sqrt{1-4A_3}\over 2}\, ,\,3\alpha-\frac 12\, ;\, z\right)
\ee
and
\be
Y_{II}=z^{3(1-\alpha)/2}F\left(1-\frac 32\alpha+{\sqrt{1-4A_3}\over
    2}\, ,\, 1-\frac 32\alpha-{\sqrt{1-4A_3}\over 2}\, ,\, {5\over 2}
-3\alpha\, ;\, z\right),
\ee
are independent.

We use the linear tranformation formula \cite{morse-feshbach},
\begin{eqnarray}
F(a,b,c;z)&=&{\Gamma(c)\Gamma(b-a)\over \Gamma(b)\gamma(c-a)}(-z)^{-a}
F\left(a,1-c+a,1-b+a;{1\over z}\right)\nonumber\\
&+&{\Gamma(c)\Gamma(a-b)\over
\Gamma(a)\Gamma(c-b)}(-z)^{-b}F\left(b,1-c+b,1-a+b;{1\over z}\right)
\end{eqnarray}
for hypergeometric functions,
where asymptotically $z\sim t$, to obtain
\be
Y\sim t^{\frac 12 +{\sqrt{1-4A_3}\over 2}},\label{atheta}
\ee
for very large times.
Subsequently, for an open universe at large times, when $R_0\sim
e^{-i\theta}\sim t$ and $\rho_0\sim t^{-3}$, 
we obtain, using (\ref{xyeqn}) and (\ref{x1}),
\bear
X&\sim& t^x\qquad ;\qquad x=-\frac 32\alpha +\frac 12\sqrt{1-4A_3},\\
\lambda &\sim&t^{-\frac 32\alpha -3+\frac 12\sqrt{1-4A_3}}.
\eear
For complex values of the square-root term
in $x$, the matter distribution becomes inhomogeneous  
only when $\alpha\le \frac {2\gamma}3$. On the other hand, for real values of
the square root, the inhomogeneity is enhanced. This occurs, for
$\gamma\approx 1$, if $\alpha\mathop{>}\limits_\sim 0.8$ or 
$\alpha\mathop{<}\limits_\sim -0.4$. Putting these results together,
we see that also for 
$\alpha\mathop{>}\limits_\sim 1$ the scale-invariant perturbation
dominates over the homogeneous background. 
The inhomogeneity grows by a factor of
$10^{3/2}$ for $\alpha\approx 1/2$ 
during the matter-dominated era (see eqn. (\ref{eta})). 

%%%%%%%%%%%%%%%%%%%
\subsection{Open universe with non-vanishing pressure \\
 $P_0\not=0\quad,\quad\alpha\not=0 $}
%%%%%%%%%%%%%%%%%%%%
\indent

In this subsection, we consider the most general equation of state for
an open universe where neither the homogeneous background
nor the perturbation contributions to the pressure vanish. 

The homogeneous density $\rho_0$, obtained by
using the equation of state $P_0=\omega\rho_0$ in the
energy-momentum tensor equation (\ref{energy1}), is
\be
\rho_0= C_1 R_0^{3(\omega-1)}
\ee
where $C_1$ is an integration constant. By substituting for $\rho_0$
in the homogeneous Einstein equations \footnote{The homogeneous
Einstein equation are simply obtained 
by using the Friedmann metric (\ref{metric}) 
in the Einstein field equation (\ref{fieldeqn}).}, we obtain
\be
{dR_0\over dt}=\sqrt{1-{8\pi G\over 3}C_1R_0^{3\omega-1}}.
\ee
The solution is the hypergeometric equation \footnote{The
expression (\ref{rreqn}) is recovered for $\omega=0$.}
\be
t={-2\sqrt C_2\over 3(\omega-1)}R_0^{{3\over 2}(1-\omega)} F\left(\frac
  12\, ,\, {3(\omega-1)\over 2( 3\omega-1)}\, ,\, 
{9\omega-5\over 2(3\omega-1)}\, ,\, -C_2R_0^{1-3\omega}\right).
\ee
Thus, since
the exact dependence of $R_0$ on $t$ is not known, the
differential equation (\ref{mastereqn}) cannot be solved exactly. 
However, asymptotically, for large
times, $R_0$ varies linearly with time and thus the differential equation
(\ref{mastereqn}) becomes
\be
\frac{d^2Y}{dt^2} +Ct^{-3\omega-3}Y
+A_3t^{-2}Y =0 ,\label{lala}
\ee
where $C$ is a constant.  Next, we substitute 
\be
Y=t^{\frac 12(1\pm\sqrt{1-4A_3})} e^{2\over (3\omega+1)\tau}F\left(
{1\over\tau}\right),
\ee  
where $t=C^{\frac 1 {1+3\omega}}\tau^{2\over 1+3\omega}$, in the differential
equation (\ref{lala}) to obtain a confluent hypergeometric 
equation \cite{morse-feshbach}.
However, asymptotically, $F(1/\tau)$ 
approaches a constant value
\cite{morse-feshbach} and we obtain, using (\ref{xyeqn}),
\be
X\sim t^{-3\alpha+\sqrt{1-4A_3}\over 2},
\ee
whose behaviour does not depend on $\omega$.
For the density
perturbation, we obtain
\be
\lambda\sim t^{\left[(3\alpha+\sqrt{1-4A_3})/2\right]-3}
\ee
and subsequently the growth parameter of the density fluctuation is
\be
{{\rm ln}(\lambda R_0^\gamma/ \rho_0)\over {\rm ln} t}
\sim t^{\left(-{3\alpha\over 2}+1+3\omega\right)}.
\ee
Thus, the fluctuation dominates over 
the background homogeneous density for all positive values of
$1+3\omega-3\alpha/2$ or for large positive values of $\alpha$ and 
$\omega\mathop{>}\limits_\sim -0.3$. 

The growth rate of the perturbation, obtained by using the above growth
parameter in equation (\ref{eta}), is much larger in 
the radiation-dominated era ($t/t_0\approx 10^8$) than in
the matter-dominated era ($t/t_0\approx 10^5$).
In the radiation-dominated period, 
the perturbation grows by a
factor of about $10^{12}$ for $\alpha\approx 1/3$ and
reduces to its lowest value at $10^4$ for $\alpha\approx 1$. 
Thus, the homogeneous Friedmann model needs to be revised 
for an open universe.

%%%%%%%%%%%%%%%%%
\section{The r\^ole of the polar pressure}
%%%%%%%%%%%%%%%%
\indent

The differential equation (\ref{mastereqn}) has been obtained by
using the equation of state, $\pi_r=\alpha\lambda$, to relate
the radial pressure to
the density via the parameter $\alpha$. So far, 
we have used the time-time component (\ref{x1})
and the radial-radial component (\ref{x2}) of the Einstein
equation. However, instead of relating the radial pressure
to the density, we could equally relate the 
polar pressure to the density
through a similar adiabatic equation of state,
\be
\pi_\theta=\alpha_\theta\lambda.
\ee
We could also use the polar-polar component of the Einstein
equation (\ref{x3}) instead of (\ref{x2}). In this section, we
show that the results obtained so far, are left intact by this
interchange.

Using equations (\ref{x1}) and (\ref{x3}) and the above 
adiabatic equation of state for the perturbations, we obtain the
the second-order differential equation 
\be
\ddot Y-\left(6\pi(\alpha_\theta-1)(\alpha_\theta\rho-P_0)-
{A_\theta}_3{k\over R_0^2}\right) Y=0 \label{mastereqntheta}
\ee
where
\be
{A_\theta}_3=(\gamma+1)\left(\alpha_\theta(\gamma-3)-\gamma+2\right)-\frac
34(\alpha_\theta-1)(3\alpha_\theta-1)=A_3+(1-\gamma).
\ee
The differential equation (\ref{mastereqntheta}) is identical to 
the previously obtained equation (\ref{mastereqn})
, with $\alpha$ and $A_3$ replaced by $\alpha_\theta$ and
 ${A_\theta}_3$, respectively. For a flat universe, $k=0$, 
this interchange is immaterial since 
the coefficient ${A_\theta}_3$ is eliminated from the 
differential equation (\ref{mastereqntheta}) and 
we obtain the same results
as those obtained in Sections IV and V.

For $k\not=0$, the differential equation is modified because of the
extra terms in ${A_\theta}_3$ as compared to $A_3$,
(\ref{a3}). However, the fractal codimension obtained from the
observational results is approximately 1 \cite{reportnero}, which
leads to the equivalence of $A_3$ and $A_{\theta 3}$. Thus, the radial
and polar pressures play the same r\^ole in 
the evolution of density fluctuations and their interchange has no
effect on our results.

%%%%%%%%%%%%%%%%%%%%%%%%%%%%%%%%%%%%%%%%%%%%%%%%%%%%%%
\section{conclusion}
%%%%%%%%%%%%%%%%%%%%%%%%%%%%%%%%%%%%%%%%%%%%%%%%%%%%%%%%
\indent

We conclude that any density fluctuation in an open universe is 
scale-invariant. In a flat universe, the scaling law is not obeyed by
density perturbations. 

We have seen that a small seed of
adiabatic inhomogeneity decays by a
factor of 100 in the matter-dominated era. We have shown that, on the
contrary, during this time, such a seed grows by a factor of $10^5$ 
in an open universe. Moreover, the 
fluctuation grows by a factor of $10^{12}$ 
in the radiation-dominated era.
Thus, we conclude that any adiabatic scale-invariant
perturbation would eventually make an open universe inhomogeneous.
On the contrary, in a flat universe, 
the homogeneity hypothesis of the Friedmann cosmology
remains valid. In a flat inflationary universe, scale-invariant
adiabatic density fluctuations grow
exponentially, leading to an inhomogeneous universe.

We also conclude that polar and radial pressures play the same r\^ole
in the dynamics of perturbation growth. Their interchange leaves our
results unaffected.

{\bf Acknowledgement \, :\,} 
We thank Conselho Federal de Desenvolvimento Cient\'\i fico e
Tecnol\'ogico (CNPq-Brazil) and Funda\c c\~ao de Amparo a Pesquisa do
Estado de S\~ao Paulo (FAPESP) for financial support.

%%%%%%%%%%%%%%%%%%%%%%%%%%%%%%%%%%%%%%%%%%%%%%%%%%%%%%%%%%%%%%%%%%%%%%%%%%%

\end{document}